\documentclass[aps,prl,twocolumn,showpacs]{revtex4}
\usepackage[dvips]{epsfig}
\usepackage{longtable}
\begin{document}
\title{On the interaction of NH($X^{3}\Sigma^{-}$) molecules with
rubidium atoms: implications for sympathetic cooling and the
formation of extremely polar molecules}
\author{Pavel Sold\'{a}n}
\author{Jeremy M. Hutson}
\affiliation{Department of Chemistry, University of Durham, South Road,
Durham, DH1~3LE, England}
\date{\today}

\begin{abstract}
The Rb-NH interaction is investigated as a prototype for
interactions between alkali metal atoms and stable molecules. For
spin-aligned Rb and NH that interact on a quartet surface
($^4A''$), the interaction is relatively weak, with a well depth
of 0.078 eV. However, there are also doublet surfaces of ion-pair
character that are very much deeper (well depth 1.372 eV) and may
be important for atom - molecule collision rates. Similar deeply
bound ion-pair states are likely to exist for other alkali atom --
molecule pairs. It is shown that a transition to a dipolar
superfluid phase will be easier to achieve for low-mass than for
high-mass species.
\end{abstract}


\maketitle

\font\smallfont=cmr7

There is great interest in the production and trapping of
translationally cold molecules, and in the possibility of
achieving quantum degeneracy in molecular gases
\cite{BG,PA,PABEC,FR}. There is special interest in dipolar
quantum gases \cite{Lewen02a}, where the trap geometry controls
the attractive or repulsive nature of the anisotropic
dipole-dipole interaction. Important goals in this area include
Bose-Einstein condensation of dipolar gases \cite{Lewen00},
realization of phase transitions of such gases in optical lattices
\cite{Lewen02b}, and formation of superfluid pairs and observation
of the Bardeen-Cooper-Schrieffer (BCS) transition to a superfluid
state in dipolar Fermi gases \cite{Gora02}. The electric dipole
moments of ultracold polar molecules could also be used as qubits
in quantum computation \cite{DeMille02}.

One very promising route for the production of cold dipolar
molecules is molecular beam deceleration \cite{Meijer03}. Meijer
and coworkers have developed a decelerator based on the molecular
Stark effect and switched inhomogeneous electric fields
\cite{Meijer99}, and have used it to slow a beam of ND$_3$
molecules and trap them at a temperature around 25 mK
\cite{Meijer00}. Beam deceleration is also applicable to other
polar molecules, such as OH and H$_2$O, but is typically only
likely to achieve temperatures of the order of 1 mK. To cool the
molecules further, towards temperatures where quantum degeneracy
might be achieved, another cooling technique is needed. One
possible route is sympathetic cooling by thermal contact with a
cold atomic gas such as $^{87}$Rb \cite{Meijer02}.

The purpose of the present paper is to investigate the potential
energy surfaces for interaction of a polar molecule with an alkali
atom, in order to help understand collisions in molecule - alkali
atom mixtures. For this prototype study, we choose NH
($X^{3}\Sigma^{-}$) interacting with with Rb ($^2S$). This system
is topical because Meijer and coworkers \cite{Meijer01} have
recently proposed a scheme for the production of ultracold NH
molecules in their ground electronic and vibrational state. In
this scheme, NH molecules in their long-lived $a^{1}\Delta$
metastable state are first slowed in a Stark decelerator. The
molecules are then excited to the $v=0$ level of the $A^3\Pi$
excited state, allowed to decay spontaneously to the $v=0$ level
of the $X^{3}\Sigma^{-}$ ground state, and finally trapped
magnetically.

An important and general point is that many molecule - alkali atom
systems have deeply bound electronic states with ion-pair
character. Alkali metal atoms have low ionization energies, while
molecules containing electronegative elements such as C, N, O and
the halogens can accept an extra electron relatively easily. There
is thus an ion-pair state (Rb$^+$ NH$^-$ in the present case) that
is asymptotically not far above the energy of the neutrals.
Because of the Coulomb attraction between the ions, the potential
surface for the ion-pair state cuts down fast in energy with
decreasing separation, and around the equilibrium geometry is
often the lowest electronic state. As will be seen below, this is
the case for RbNH.

The $X^{3}\Sigma^{-}$ ground state of NH has 2 electrons with
parallel spins in $\pi$ antibonding orbitals. At collinear
geometries (with $C_{\infty v}$ point group symmetry), there are
two electronic states of RbNH, $^{4}\Sigma^{-}$ and
$^{2}\Sigma^{-}$, which correlate with the NH($X^{3}\Sigma^{-}$) +
Rb($^{2}S$) dissociation limit. At nonlinear geometries (with
$C_{s}$ symmetry), these two states become $^{4}A''$ and
$^{2}A''$. These states are principally bound by dispersion
forces. However, as mentioned above, there is also an ion-pair
threshold Rb$^+$($^{1}S$) + NH$^-$($^{2}\Pi$) less than 4 eV above
the neutral threshold. The resulting $^{2}\Pi$ ion-pair state of
RbNH is subject to the Renner-Teller effect and splits at
nonlinear geometries into two states with the unpaired electron
(hole) either in the triatomic plane ($^{2}A'$) or perpendicular
to it ($^{2}A''$). There are thus two $^{2}A''$ states, one of
ion-pair character and one dispersion-bound, which form a conical
intersection at linear geometries, where one has $^{2}\Pi$ and the
other $^{2}\Sigma^{-}$ symmetry. [This is a conical intersection
on the two-dimensional surface obtained when the NH bond length
$r_{\rm NH}$ is fixed; in three dimensions, the two surfaces
intersect on a seam at linear geometries.]

The influence of ion-pair states on collisions of alkali metal
atoms and similar species has been called the ``harpoon" mechanism
\cite{L&B}. If the electron transfer takes place at long range, as
is often the case, large collision cross sections result. The
range of the harpooning is determined by the position $R_{\rm X}$
of the crossing point (conical intersection). This can be
estimated from
\begin{equation}
\label{Eq1} \Delta E_{0} = \frac{e^{2}}{4\pi\varepsilon_{0}R_{\rm
X}},
\end{equation}
where $\Delta E_{0}$ is the difference between the Rb ionization
energy and the NH$^{-}$ vertical ionization energy (NH electron
affinity). The experimental values for the Rb ionization potential
and the NH vertical electron affinity are 4.177 eV \cite{Moore}
and 0.387 eV respectively; the latter is obtained from the
adiabatic NH electron affinity (0.381 eV) \cite{EA} and a
calculated value of the difference between the zero-point energies
of NH and NH$^{-}$ (0.006 eV) \cite{ZPE}. This gives an estimate
of $R_{\rm X}\approx 3.79$~\AA. Because in NH$^{-}(^{2}\Pi)$ the
negative charge is concentrated on the nitrogen atom, $R_{\rm X}$
represents the distance between N and Rb.

The nature of the dispersion-bound and ion-pair states depends on
the electron configuration of the molecular partner. For Rb-OH,
for example, the dispersion-bound states are $^{1}\Pi$ and
$^{3}\Pi$ at linear geometries, splitting into $^{1}A'$,
$^{1}A''$, $^{3}A'$ and $^{3}A''$ at non-linear geometries, while
the ion-pair state is a closed-shell singlet ($^{1}\Sigma^{+}$ or
$^{1}A'$). As for Rb-NH, therefore, there is no ion-pair state for
the highest allowed spin multiplicity. For Rb-CH, the
dispersion-bound states are the same but there are many more
ion-pair states arising from a $\pi^{2}$ configuration:
$^{3}\Sigma^{-}$ ($^{3}A''$), $^{1}\Delta$ ($^{1}A'$ and
$^{1}A''$) and $^{1}\Sigma^{+}$ ($^{1}A'$). The only
dispersion-bound state that does not have an ion-pair counterpart
in this case is the $^{3}A'$ state.

The high values of the dipole polarizabilities for alkali-metal
atoms imply large dispersion forces. Because of this, the crossing
point is in the classically allowed region of the dispersion-bound
state for many molecule - alkali atom systems. This is the case
for RbNH. The ``harpooning" is thus a barrierless process and no
tunnelling is required for the charge transfer.

\begin{figure}[t]
\begin{center}
\epsfig{file=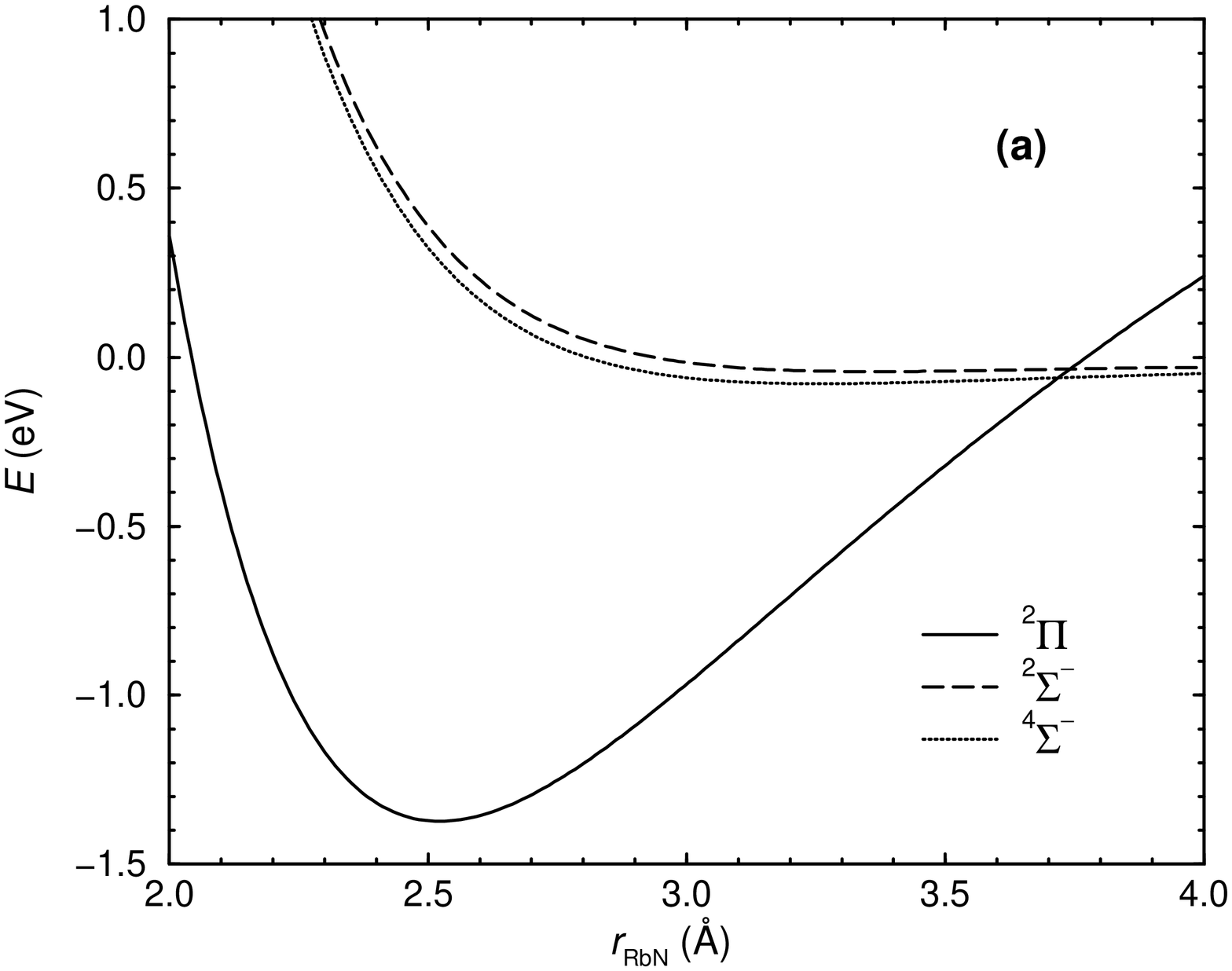,angle=-90,width=85mm}
\epsfig{file=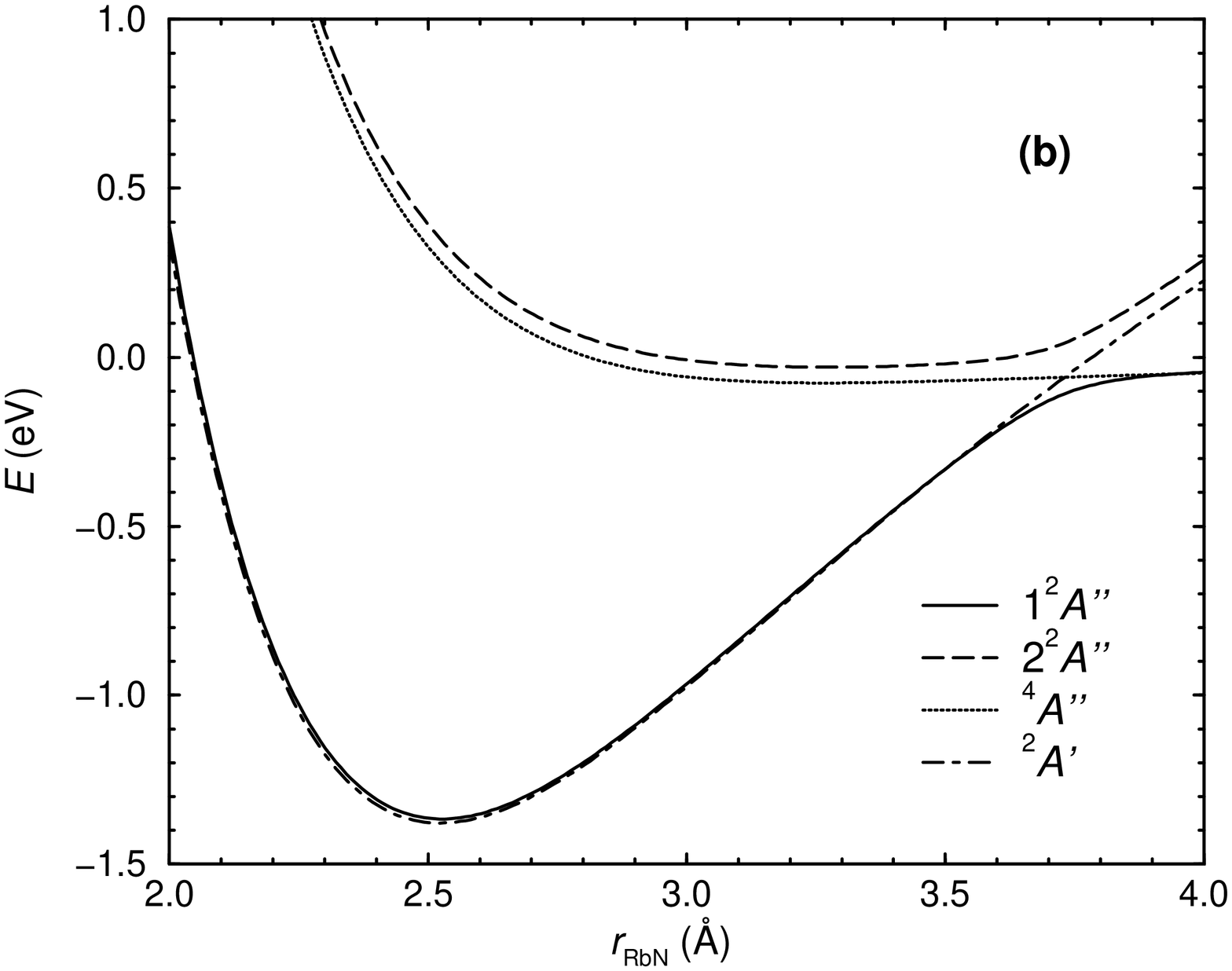,angle=-90,width=85mm}
\end{center}
\caption{CASSCF(10,3)/MRCI potential energy curves of RbNH with
fixed $r_{\rm NH}=1.0308$ \AA\ and fixed Rb-N-H angle of
180$^{\circ}$ (a) and 170$^{\circ}$ (b). (a) At linear geometries
the crossing point is at 3.77 \AA. (b) At non-linear geometries
the crossing is avoided.} \label{fig1}
\end{figure}

To obtain a more quantitative picture of the RbNH electronic
states, we have carried out {\em ab initio} calculations at the
CASSCF(10,3)/MRCI level (complete active space self-consistent
field / multireference configuration interaction). For N and H, we
used the aug-cc-pVTZ basis sets of Dunning \cite{Dunning} in
uncontracted form, and for Rb we used the quasirelativistic
small-core ECP10MWB effective core potential of Leininger
\textit{et al.} \cite{ECP}, augmented by the valence basis set
from Ref.\ \cite{Soldan03}. The (10,3) active space covers all the
occupied orbitals on N and H and all the large-valence and 5$p$
orbitals on Rb. Only the electrons in the 1$s$ orbital on N were
excluded from the correlation calculations. All the calculations
were performed using the MOLPRO package \cite{MOLPRO}.

Figure \ref{fig1} shows the resulting curves for Rb-N-H angles of
180$^{\circ}$ and 170$^{\circ}$ (for fixed NH bond length). The
minimum on the quartet surface is at a depth of 0.078 eV at
$r_{\rm RbN} =3.26$~\AA, while that on the doublet ion-pair
surface is at a depth of 1.372 eV at $r_{\rm RbN}=2.53$~\AA. At
linear geometries the $^{2}\Pi$ and $^{2}\Sigma^{-}$ curves cross
at $R_{\rm X}=3.77$~\AA. At non-linear geometries the crossing is
replaced by an avoided crossing. Note that the minimum of the
$^{2}\Sigma^{-}$ curve is at a shorter Rb-N distance (around
3.4~\AA) than the crossing point, so that at 170$^{\circ}$ the
$1^{2}A''$ curve has just one minimum.

The collisional processes in Rb-NH mixtures will depend greatly on
the magnetic states that are present. If both
NH($X^{3}\Sigma^{-}$) and Rb($^{2}S$) are in spin-stretched states
($|M_{F}|=F=F_{\rm max}$), collisions will take place on the
$^{4}A''$ surface. If the NH($X^{3}\Sigma^{-}$) molecules are
initially in their ground rovibrational state ($v=0,N=0,J=1,M=1$)
\cite{Meijer01}, only elastic and reorientation (spin-changing)
atom-molecule collisions can occur. Under these circumstances,
trap loss will be caused only by spin-changing collisions and by
three-body recombination processes.

If the species present are not in spin-stretched states,
collisions will also involve the doublet surfaces. In this case,
the harpoon mechanism may be important and may enhance the rates
of reorientation collisions. In addition, the transient collision
complexes formed during the collision will be highly polar, and
may be expected to have enhanced interactions with other species.
Three-body collision rates may thus be large. Collisions with
third bodies might also result in vibrational quenching and
formation of bound RbNH($^{2}\Pi)$ molecules with ionic character,
and consequently in trap loss. However, because of the short-range
nature of this type of three-body recombination (which can occur
only when the distance between the N and Rb is smaller than
$R_{\rm X}$), it cannot be described in terms of the
fragment-fragment scattering lengths and a three-body parameter.

On the other hand, the harpoon mechanism followed by, say,
photoassociation offers a possible means for production of
extremely polar molecules, with electric dipole moments well over
5~D (the value of the electric dipole moment of RbNH($^{2}\Pi$)
varies from 8.7~D to 15.3~D between the potential minimum and the
crossing point). Such molecules might provide a valuable probe of
low-temperature physics of dipolar gases, because phenomena
related to dipole-dipole interactions, such as linking of polar
molecules \cite{Bohn03}, will be significantly enhanced.

The possibility of producing ultracold fermionic molecules and
achieving a transition to a superfluid phase is particularly
interesting. The molecule - alkali atom complexes have
particularly large dipole moments, so at first sight are very
promising candidates, but it is worthwhile to seek a more
quantitative characterization of which dilute dipolar Fermi gases
will be most suitable for transitions to the superfluid phase.

Following Baranov \textit{et al.}\ \cite{Gora02}, we assume that
the mean dipole-dipole energy per particle is much smaller than
the Fermi energy,
\begin{equation}
n d^{2}\ll \varepsilon_{F}, \label{Eq2}
\end{equation}
where $n$ is the number density of the gas, $d$ is the molecular
electric dipole moment and
$\varepsilon_{F}$ is the Fermi energy.
Because the relation ``much smaller than" is rather vague and
difficult to work with, we introduce an upper bound $\alpha_{\rm
UCL}$ for the ratio of $n d^2$ to $\varepsilon_{F}$
\begin{equation}
\frac{n d^{2}}{\varepsilon_{F}}\le \alpha_{\rm UCL} \ll 1.
\label{Eq2b}
\end{equation}
Since there appear to be severe technical limitations to the
cooling of atomic Fermi gases below 0.2$\varepsilon_{F}/k_{B}$
\cite{Jin01} (where $k_{B}$ is Boltzmann's constant), we introduce
a lower bound for the ratio of the critical temperature for the
BCS transition, $T_{c}$, to the Fermi temperature
$\varepsilon_{F}/k_{B}$,
\begin{equation}
\frac{k_{B} T_{c}}{\varepsilon_{F}} \geq \beta_{\rm BCS}.
\label{Eq3}
\end{equation}
Baranov \textit{et al.}\ \cite{Gora02} have derived an analytical
expression for this ratio,
\begin{equation}
\frac{k_{B} T_{c}}{\varepsilon_{F}}=1.44\exp{\left\{-\frac{\pi
\hbar}{2 p_{F} |a_{d}|}\right\}}, \label{Eq3a}
\end{equation}
where $p_{F}=\hbar (6 \pi^{2} n)^{1/3}$ is the Fermi momentum and
$a_{d}=-{2 m d^{2}}/(\pi^{2} \hbar^{2})$ is the effective
scattering length.

We can combine the above expressions and  derive lower and upper
bounds for the product $m d^{2} n^{1/3}$,
\begin{equation}
\frac{\pi^{3} \hbar^{2}}{4 (6\pi^{2})^{1/3}
\ln{\left(\frac{1.44}{\beta_{\rm BCS}}\right)}} \leq m d^{2}
n^{1/3} \leq \frac{1}{2} \hbar^{2} (6\pi^{2})^{2/3} \alpha_{\rm
UCL}. \label{Eq4}
\end{equation}
For a given species ({\em i.e}, given $m$ and $d$), Eq.\
(\ref{Eq4}) gives the condition for the number density at which
the dipolar Fermi gas reaches the so-called ultracold limit, Eq.\
(\ref{Eq2b}), and the transition temperature satisfies Eq.\
(\ref{Eq3}). Obviously densities closer to the upper bound would
give a higher Fermi temperature (and consequently higher critical
temperature) and might be preferable from an experimental point of
view. If for given $\alpha_{\rm UCL}$ we introduce maximal density
$n_{\rm max}$  by the equation
\begin{equation}
m d^{2} n_{\rm max}^{1/3}=\frac{1}{2}\hbar^{2} (6\pi^{2})^{2/3}
\alpha_{\rm UCL}, \label{Eq5}
\end{equation}
then the ratio from Eq.\ (\ref{Eq3}) is reduced to
\begin{equation}
\frac{k_{B} T_{c}}{\varepsilon_{F}}=1.44\exp{\left\{-\frac{\pi}{12
\alpha_{\rm UCL}}\right\}}. \label{Eq6}
\end{equation}
Values of $\alpha_{\rm UCL}$ between 0.1 and 0.2 give $k_{B}
T_{\rm c}/\varepsilon_{\rm F}$ ratios between 0.10 and 0.39.

Fermionic $^{14}$NH has an electric dipole moment of $d=1.389$ D
\cite{NHdip}, which gives a rather large absolute value of the
effective scattering length, $a_{d}=-876$ \AA. Setting
$\alpha_{\rm UCL}=0.1$ leads to a ``reasonable'' density, $n_{\rm
max}=5.4\times10^{12}$ cm$^{-3}$. Consequently, we obtain a very
favorable Fermi temperature $\varepsilon_{F}/k_B=757$ nK and
critical temperature $T_{c}=80$ nK.

In the case of $^{87}$Rb$^{14}$NH($^{2}\Pi$), the large dipole
moment and large mass give an enormous absolute value of the
effective scattering length, $a_{d}=-234487$ \AA. Setting
$\alpha_{\rm UCL}=0.2$ gives a very low density, $n_{\rm
max}=2.3\times10^{6}$, which combined with the large mass gives
tiny values, 6.2 pK and 2.4 pK, for the Fermi temperature and
critical temperature respectively. That means that, despite the
very large dipole moment, it would be experimentally difficult to
achieve the transition to a superfluid phase in a dilute gas of
$^{87}$Rb$^{14}$NH.

\begin{table*}[t]
\caption{Dipole moment $d$ (D), effective scattering length
$a_{d}$ (\AA), maximal density $n_{\rm max}$ (cm$^{-3}$), Fermi
temperature $\varepsilon_{F}/k_B$ (nK) and critical temperature
$T_{c}$ (nK) for selected fermionic molecules in ground electronic
states.} \label{table1}
\begin{ruledtabular}
\begin{tabular}{rlrlrrlrr}
& & & \multicolumn{3}{c}{$\alpha_{\rm UCL}=0.1$} & \multicolumn{3}{c}{$\alpha_{\rm UCL}=0.2$} \\
\cline{4-6} \cline{7-9}
 & $d$ & $a_{d}$ & $n_{\rm max}$ & $\varepsilon_{F}/k_B$ & $T_{c}$ & $n_{\rm max}$ & $\varepsilon_{F}/k_B$ & $T_{c}$ \\
\colrule
$^{6}$Li$^{23}$Na & 0.46\footnote{Reference\ \cite{LiNa}} & -186 &
5.7$\times$10$^{14}$ & 8.7$\times$10$^{3}$ &
917 & 4.6$\times$10$^{15}$ & 3.5$\times$10${^4}$ & 1.4$\times$10$^{4}$ \\
$^{14}$NH & 1.389\footnote{Reference\ \cite{NHdip}} & -876 & 5.4$\times$10$^{12}$ & 757 & 80 & 4.3$\times$10$^{13}$ & 3030 & 1180 \\
$^{15}$ND$_{3}$ & 1.5\footnote{Reference\ \cite{ND3}} & -1432 & 1.2$\times$10$^{12}$ & 202 & 21 & 9.9$\times$10$^{12}$ & 809 & 315 \\
H$^{12}$C$^{14}$N & 2.98\footnote{Reference\ \cite{HCN}} & -7258 & 9.5$\times$10$^{9}$ & 6.1 & 0.65 & 7.6$\times$10$^{10}$ & 25 & 9.6 \\
$^{7}$Li$^{40}$K & 3.41\footnotemark[1] & -16919 & 7.5$\times$10$^{8}$ & 0.65 & 0.068 & 6.0$\times$10$^{9}$ & 2.6 & 1.0 \\
$^{7}$Li$^{14}$NH & 5.1\footnote{Present work; RHF value at
RCCSD(T)/cc-pV5Z optimized linear geometry $r_{\rm NH}=1.014$~\AA\
and $r_{\rm LiN}=1.695$~\ \AA.} & -17335 &
7.0$\times$10$^{8}$ & 1.32 & 0.139 & 5.6$\times$10$^{9}$ & 5.3 & 2.1 \\
$^{87}$Rb$^{14}$NH & 8.72\footnote{Present work; CASSCF(10,3)/MRCI
value, see text for details.} & -234487 & 2.8$\times$10$^{5}$ &
0.0016 & 0.0002 & 2.3$\times$10$^{6}$ & 0.0062 & 0.0024 \\
\end{tabular}
\end{ruledtabular}
\end{table*}

The above discussion demonstrates that the combination of large
mass {\em and} large electric dipole moment is undesirable for
production of superfluid dipolar Fermi gases. Large mass implies
low Fermi temperature, which consequently makes the superfluid
transition temperature very low. Table \ref{table1} summarizes the
situation for several fermionic molecules that are of interest to
the ultracold community. The maximum number densities, Fermi
temperatures and critical temperatures are listed for two values
of $\alpha_{\rm UCL}$, 0.1 and 0.2. Table \ref{table1} illustrates
that the dipole moment and molecular mass are {\em both} relevant
characteristics when considering transitions to superfluid phases
in dipolar Fermi gases. Lighter fermionic molecules with a
reasonably large dipole moment, such as $^{14}$NH or
$^{6}$Li$^{23}$Na, are more suitable for transition-to-superfluid
experiments than heavy molecules with very large dipole moments.


\begin{thebibliography}{99}

\bibitem{BG} J.\,D.\ Weinstein \textit{et al.},
{Nature (London)} \textbf{395}, 148 (1998).

\bibitem{PA} A.\ Fioretti \textit{et al.},
{Phys.\ Rev.\ Lett.} \textbf{80}, 4402 (1998); A.\,N.\ Nikolov
\textit{et al.}, \textit{ibid.} \textbf{82}, 703 (1999); C.\
Gabanini \textit{et al.}, \textit{ibid.} \textbf{84}, 2814 (2000).

\bibitem{PABEC}
R.\ Wynar \textit{et al.}, {Science} {\bf 287}, 1016 (2000);
J.\,M.\ Gerton, D.\ Strekalov, I.\ Prodan, and  R.\,G.\ Hulet,
{Nature (London)} \textbf{408}, 692 (2000); C.\ McKenzie
\textit{et al.}, {Phys.\ Rev.\ Lett.} {\bf 88}, 120403 (2002).

\bibitem{FR} E.\,A.\ Donley, N.\,R.\ Claussen, S.\,T.\ Thompson, and
C.\,E. Wieman, {Nature (London)} \textbf{417}, 529 (2002);
P.\ Zoller,
\textit{ibid.} \textbf{417}, 493 (2002).

\bibitem{Lewen02a}
M.\ A.\ Baranov \textit{et al.},
{Physica Scripta} \textbf{T102}, 74 (2002).


\bibitem{Lewen00}
L.\ Santos, G.\,V.\ Shlyapnikov, P.\ Zoller, and M.\ Lewenstein,
{Phys.\ Rev.\ Lett.} \textbf{85}, 1791 (2000); \textit{ibid.}
\textbf{88}, 139904(E) (2002).

\bibitem{Lewen02b}
K.\ G\'{o}ral, L.\ Santos, and M.\ Lewenstein,
\textit{ibid.} \textbf{88}, 170406 (2002);
B.\ Damski \textit{et al.}
\textit{ibid.} \textbf{90}, 110401 (2003).

\bibitem{Gora02}
M.\,A.\ Baranov, M.\,S.\ Mar'enko, Val.\,S.\ Rychkov, and G.\,V.\
Shlyapnikov,
{Phys.\ Rev.\ A} \textbf{66}, 013606 (2002).

\bibitem{DeMille02} D.\ DeMille,
{Phys.\ Rev.\ Lett.} \textbf{88}, 067901 (2002).

\bibitem{Meijer03}
H.\,L.\ Bethlem and G.\ Meijer,
{Int.\ Rev.\ Phys.\ Chem.} \textbf{22}, 73 (2003).

\bibitem{Meijer99} {H.\,L.\ Bethlem, G.\ Berden, and G.\ Meijer},
{Phys.\ Rev.\ Lett.} \textbf{83}, 1558 (1999);
H.\,L.\ Bethlem \textit{et al.},
\textit{ibid.} \textbf{84}, 5744 (2000).

\bibitem{Meijer00}
H.\,L.\ Bethlem \textit{et al.}, {Nature (London)} \textbf{406},
491 (2000).

\bibitem{Meijer02} G.\ Meijer,
in \textit{Interactions of Cold Atoms and Molecules}, edited by
P.\ Sold\'{a}n, M.\,T.\ Cvita\v{s}, J.\,M.\ Hutson, and C.\,S.\
Adams (CCP6, Daresbury, 2002), p.\ 35.

\bibitem{Meijer01} S.\,Y.\,T.\ van de Meerakker \textit{et al.},
{Phys.\ Rev.\ A} \textbf{64}, 041401 (2001).

\bibitem{L&B} R.\,D.\ Levine and R.\,B.\ Bernstein,
\textit{Molecular Reaction Dynamics and Chemical Reactivity}
(Oxford University Press, New York, 1987), p.\ 134.

\bibitem{Moore} C.\,E.\ Moore, \textit{Atomic Energy Levels},
(National Bureau of Standards, 1952), Vol.\ II.

\bibitem{EA} R.\,J.\ Celotta, R.\,A.\ Bennett, and J.\,L.\ Hall,
J.\ Chem.\ Phys.\ \textbf{60}, 1740 (1974);
P.\,C.\ Engelking and W.\,C.\ Lineberger,
\textit{ibid.} \textbf{65}, 4323 (1976).

\bibitem{ZPE} G.\,L.\ Gutsev and R.\,J.\ Bartlett,
Chem.\ Phys.\ Lett.\ \textbf{265}, 12 (1997).

\bibitem{Dunning} T.\,H.\ Dunning, Jr.,
{J.\ Chem.\ Phys.} \textbf{90}, 1007 (1989).

\bibitem{ECP} T.\ Leininger \textit{et al.},
{Chem.\ Phys.\ Lett.} \textbf{255}, 274 (1996).

\bibitem{Soldan03} P.\ Sold\'{a}n, M.\,T.\ Cvita\v{s}, and J.\,M.\
Hutson,
{Phys.\ Rev.\ A} \textbf{67}, 054702 (2003).

\bibitem{MOLPRO}
MOLPRO is a package of {\it ab initio} programs written by H.-J.
Werner and P.\,J. Knowles with contributions from others; for more
information see the www page http://www.tc.bham.ac.uk/molpro/

\bibitem{Bohn03} A.\,V.\ Avdeenkov and J.\,L.\ Bohn,
{Phys.\ Rev.\ Lett.} \textbf{90}, 043006 (2003).

\bibitem{Jin01} B.\ DeMarco, S.\,B.\ Papp, and D.\,S.\ Jin,
{Phys.\ Rev.\ Lett.} \textbf{86}, 5409 (2001).

\bibitem{NHdip} Rajendra Pd.\ and P.\ Chandra,
{J.\ Chem.\ Phys.} \textbf{114}, 7450 (2001).

\bibitem{LiNa} V.\,J.\ Landolt-B\"{o}rnstein, \textit{Molecular
Constants} (Springer-Verlag, Berlin, 1974), Vol.\ II/6.

\bibitem{ND3} A.\ Halkier and P.\,R.\ Taylor,
{Chem.\ Phys.\ Lett.} \textbf{285}, 133 (1998).

\bibitem{HCN} G.\ Maroulis and C.\ Pouchan,
{Theor.\ Chim.\ Acta} \textbf{93}, 131 (1996).

\end{thebibliography}
\end{document}